\documentclass[10pt]{article}
\usepackage{fullpage,citesort,epsfig,graphics,amsbsy}
\usepackage{psfrag}
\usepackage[normal,small]{caption}
\newcommand{\beq}{\begin{equation}}
\newcommand{\eeq}{\end{equation}}
\newcommand{\bea}{\begin{eqnarray}}
\newcommand{\eea}{\end{eqnarray}}

\newcommand{\Tr}{{\rm Tr}}
\newcommand{\be}{\begin{equation}}
\newcommand{\ee}{\end{equation}}
\newcommand{\bq}{\begin{eqnarray}}
\newcommand{\eq}{\end{eqnarray}}

\def\math{\mathsurround=0pt }
\def\leftrightarrowfill{$\math \mathord\leftarrow \mkern-6mu \cleaders\hbox{$\mkern-2mu \mathord- \mkern-2mu$}\hfill
 \mkern-6mu \mathord\rightarrow$}
\def\overleftrightarrow#1{\vbox{\ialign{##\crcr
     \leftrightarrowfill\crcr\noalign{\kern-1pt\nointerlineskip}
     $\hfil\displaystyle{#1}\hfil$\crcr}}}

\newcommand{\bfs}{\boldsymbol}

\let\l=\lambda

 \def\bd{\begin{document}} \def\ed{\end{document}}
\def\ds{\documentstyle} \let\fr=\frac \let\bl=\bigl \let\br=\bigr
\let\Br=\Bigr \let\Bl=\Bigl
\let\bm=\bibitem
\let\na=\nabla
\let\pa=\partial \let\ov=\overline
\def\ft#1#2{{\textstyle{{\scriptstyle #1}\over {\scriptstyle #2}}}}
\def\fft#1#2{{#1 \over #2}}
\def\vp{\varphi}
\def\sst#1{{\scriptscriptstyle #1}}
\def\oneone{\rlap 1\mkern4mu{\rm l}}
\def\td{\tilde}
\def\wtd{\widetilde}
\def\dalemb#1#2{{\vbox{\hrule height .#2pt
        \hbox{\vrule width.#2pt height#1pt \kern#1pt
                \vrule width.#2pt}
        \hrule height.#2pt}}}
\def\square{\mathord{\dalemb{6.8}{7}\hbox{\hskip1pt}}}
\def\wtd{\widetilde}
\def\R{\rlap{\rm I}\mkern3mu{\rm R}}
\def\im{{\rm i}}
\def\tilg{\tilde{g}}
\def\tilF{\tilde{F}}
\def\tilA{\tilde{A}}
\def\varf{\varphi}
\def\tilf{\tilde{\phi}}
\def\tilh{\tilde{h}}
\def\rme{{\rm e}}
\def\ep{\epsilon}
\def\0{{(0)}}
\def\9{{(9)}}
\def\8{{(8)}}
\def\7{{(7)}}
\def\6{{(6)}}
\def\5{{(5)}}
\def\4{{(4)}}
\def\3{{(3)}}
\def\2{{(2)}}
\def\1{{(1)}}
\newcommand{\trace}{{\rm Tr}}
\newcommand{\ub}{\overline{U}}
\newcommand{\vb}{\overline{V}}
\newcommand{\uh}{\widehat{U}}
\newcommand{\vh}{\widehat{V}}
\newcommand{\ubh}{\overline{\widehat{U}}}
\newcommand{\vbh}{\overline{\widehat{V}}}
\newcommand{\lb}{\bar{\l}}
\newcommand{\Fb}{\overline{F}}
\newcommand{\Fh}{\widehat{F}}
\newcommand{\Fbh}{\overline{\widehat{F}}}
\newcommand{\Ab}{\overline{A}}
\newcommand{\Ah}{\widehat{A}}
\newcommand{\Abh}{\overline{\widehat{A}}}
\newcommand{\Gb}{\overline{G}}
\newcommand{\Gh}{\widehat{G}}
\newcommand{\Gbh}{\overline{\widehat{G}}}
\newcommand{\Pb}{\overline{P}}
\newcommand{\Ph}{\widehat{P}}
\newcommand{\Pbh}{\overline{\widehat{P}}}
\newcommand{\Qb}{\overline{Q}}
\newcommand{\Qh}{\widehat{Q}}
\newcommand{\Qbh}{\overline{\widehat{Q}}}
\newcommand{\Bb}{\overline{B}}
\newcommand{\Bh}{\widehat{B}}
\newcommand{\Bbh}{\overline{\widehat{B}}}
\newcommand{\fhns}{\hat{F}^{\rm (NS)}}
\newcommand{\fhrr}{\hat{F}^{\rm (RR)}}
\newcommand{\ahns}{\hat{A}^{\rm (NS)}}
\newcommand{\ahrr}{\hat{A}^{\rm (RR)}}
\newcommand{\hhrr}{\hat{H}^{\rm (RR)}}
\newcommand{\hchi}{\hat{\chi}}
\newcommand{\hphi}{\hat{\phi}}
\newcommand{\htau}{\hat{\tau}}
\newcommand{\cG}{{\cal G}}
\newcommand{\cGb}{\overline{{\cal G}}}
\newcommand{\cH}{{\cal H}}
\newcommand{\cP}{{\cal P}}
\newcommand{\cPb}{\overline{{\cal P}}}
\newcommand{\cQ}{{\cal Q}}
\newcommand{\cQb}{\overline{{\cal Q}}}
\newcommand{\cM}{{\cal M}}
\newcommand{\cN}{{\cal N}}
\newcommand{\cO}{{\cal O}}
\newcommand{\cD}{{\cal D}}
\newcommand{\cL}{{\cal L}}
\newcommand{\cA}{{\cal A}}
\newcommand{\cB}{{\cal B}}
\newcommand{\hg}{\hat{g}}
\newcommand{\cE}{{\cal E}}

\newcommand{\vpp}{\mbox{$\langle{\scriptstyle++}\rangle$}}
\newcommand{\vmp}{\mbox{$\langle{\scriptstyle-+}\rangle$}}
\newcommand{\vppp}{\mbox{$\langle{\scriptstyle+++}\rangle$}}
\newcommand{\vmpp}{\mbox{$\langle{\scriptstyle-++}\rangle$}}
\newcommand{\vpmp}{\mbox{$\langle{\scriptstyle+-+}\rangle$}}

\begin{document}
\setlength{\captionmargin}{20pt}
\begin{titlepage}
\begin{flushright}
UFIFT-HEP-03-26\\
hep-th/0311026
\end{flushright}

\vskip 3cm

\begin{center}
\begin{Large}
{\bf Fields in the Language of String:\\
Divergences and Renormalization
\footnote{Supported 
in part by the Department
of Energy under Grant No. DE-FG02-97ER-41029. 
}}
\end{Large}

\vskip 2cm
{\large 
 Charles B. Thorn\footnote{E-mail  address: {\tt thorn@phys.ufl.edu}}
}
\vskip0.20cm
{\it Institute for Fundamental Theory\\
Department of Physics, University of Florida,
Gainesville FL 32611}


\vskip 1.0cm
\end{center}

\begin{abstract}\noindent
This paper describes my talk given to the 27th Johns Hopkins
Workshop: Symmetries and Mysteries of M Theory, G\"oteborg, Sweden,
24-26 August, 2003.
After a brief introduction to the lightcone 
worldsheet formalism \cite{bardakcit}
for summing the planar diagrams of field theory, I explain how
the {\it uv} divergences of quantum field theory translate to the
new language of string. It is shown through one loop that, 
at least for scalar cubic vertices,
the counter-terms necessary for Poincar\'e invariance
in space-time dimensions $D\leq 6$ are indeed
local on the worldsheet. The extension to cover the 
case of gauge field vertices will be more complicated due to the
extra divergences at $p^+=0$ in lightcone gauge. 
\end{abstract}
\vfill
\end{titlepage}
\section{Introduction}
Two years ago Bardakci and I developed a new formalism
\cite{bardakcit}
for mapping the sum of all planar diagrams \cite{thooftlargen} of a cubic
scalar quantum field theory onto a two dimensional
system defined on the worldsheet of lightcone string theory. 
Since then the formalism has been extended
to cover Yang-Mills theory \cite{thornsheet} and
extended supersymmetric gauge theories with ${\cal N}=1,2,4$
\cite{gudmundssontt}.

Much of my talk to this workshop was devoted to a
pedagogical explanation of the new worldsheet formalism.
However, since this part of the talk was virtually
the same as one given at the August 2003 lightcone meeting
\cite{thorndurham}, I limit this introduction to a brief
synopsis of the
formulas needed to understand the new results reported
here on how
field theoretic divergences can be dealt with
{\it locally} on the worldsheet.
The worldsheet is based on light-cone
parameters, an imaginary time $\tau=ix^+=i(t+z)/\sqrt2$
in the range $0\leq\tau\leq T$,
and a worldsheet spatial coordinate $0\leq\sigma\leq p^+$
chosen so that the $p^+$ density is uniform \cite{goddardgrt}. 

The key to the worldsheet representation of 
an arbitrary planar diagram is that of
a free scalar gluon propagator, 
$\theta(T)e^{-T({\boldsymbol p}^2+\mu^2)/2p^+}$
in lightcone variables. It is based on the
remarkable identity \cite{bardakcit}
\bea
\exp\left\{-{T{\boldsymbol p}^2\over2p^+}\right\}
&=&\int DcDbD{\boldsymbol q}\exp\left\{-
\int d\tau \int_0^{p^+}d\sigma\left[{1\over2}{\bfs q}^{\prime2}
-{\bfs b}^\prime{\bfs c}^\prime\right]\right\}
\eea
where Dirichlet boundary conditions are imposed
$\dot{{\boldsymbol q}}=0$ at $\sigma=0,p^+$, and
also ${\boldsymbol q}(p^+)-{\boldsymbol q}(0)={\boldsymbol p}$.
The Grassmann variables,
with boundary conditions ${\boldsymbol b}={\boldsymbol c}=0$
at $\sigma=0,p^+$ assure the correct measure and
${}^\prime$ is shorthand for $\partial/\partial\sigma$. The
absence of time derivatives in $S$ reflects the topological nature
of the free worldsheet dynamics.
Note that in $D=d+2$ space-time dimensions we have all together
$d/2$ sets of ${\boldsymbol b}, {\boldsymbol c}$ ghost pairs, 
denoted by bold-faced letters.

We give rigorous meaning to this formula
using a worldsheet lattice \cite{gilest}: 
$\tau=ka$ and $\sigma=lm$ with $T=Na$ and $P^+=Mm$. The limit
of a continuous worldsheet is equivalent to the
double limit $M,N\to\infty$ with $N/M=(T/P^+)(m/a)$
fixed. Then defining
\bea
S&\equiv&{a\over2m}\sum_{j=1}^N\sum_{i=0}^{M-1}
({\boldsymbol q}^j_{i+1}-{\boldsymbol q}^j_i)^2
-{a\over m}\sum_{j=1}^N
\left[(1+\rho){\boldsymbol b}^j_1{\boldsymbol c}^j_1
+{\boldsymbol b}^j_{M-1}{\boldsymbol c}^j_{M-1}
+\sum_{i=1}^{M-2}({\boldsymbol b}^j_{i+1}
-{\boldsymbol b}^j_i)({\boldsymbol c}^j_{i+1}-{\boldsymbol c}^j_i)\right]
\label{freeghostaction}\\
&\equiv& S_q + S_g, \nonumber
\eea
the master formula on the worldsheet lattice is \cite{bardakcit}
\begin{eqnarray}
\left(1-{a\mu^2\over dmM}\right)^{dN/2}
\exp\left\{-{Na{\boldsymbol p}^2\over2mM}\right\}
&=&{1\over(1+\rho)^{dN/2}}\int\prod_{j=1}^N\prod_{i=1}^{M-1} 
{d{\boldsymbol c}^j_id{\boldsymbol b}^j_i\over2\pi} 
d{\boldsymbol q}^j_i\ e^{-S},
\label{nstepbits}
\end{eqnarray}
with boundary conditions ${\boldsymbol q}_0^j
={\boldsymbol q}_0$, ${\boldsymbol q}_M^j
={\boldsymbol q}_0+{\boldsymbol p}$, ${\boldsymbol b}_{0,M}^j
={\boldsymbol c}_{0,M}^j=0$. 
The parameter $\rho=\mu^2a/(dm-\mu^2a)$
provides a mass $\mu$ for the gluon in the continuum limit.
The prefactor on the right can be associated with the
left boundary.

The worldsheet lattice provides a template for summing
all planar diagrams in the cubic theory.
We can use the ratio of lattice constants
$m/a$, with units of energy/time, to define a dimensionless coupling 
\bea
{\hat g}^2\equiv{g^2\over64\pi^3}\left({m\over 2\pi a}\right)^{(d-4)/2}.
\eea
The worldsheet for the general planar diagram has
an arbitrary number of vertical solid lines marking the
location of the internal boundaries corresponding
to loops. Each interior link $j,j-1$ of
a solid line at spatial location $k$ requires a factor of
$\delta({\boldsymbol q}_{k}^j-{\boldsymbol q}_{k}^{j-1})$.
To supply such factors, assign an Ising spin $s_k^j
=\pm1$ to each site of the lattice. We assign $+1$ if the site $(k,j)$ 
is crossed by a vertical solid line, $-1$ otherwise. 
We also use the spin up projector $P_k^j=(1+s_k^j)/2$.
We implement the Dirichlet
conditions on boundaries using the Gaussian representation of
the delta function:
\begin{eqnarray}
\left({2\pi m\over a}\right)^{d/2}\delta({\boldsymbol q}_{i}^j
-{\boldsymbol q}_{i}^{j-1})
&=&\lim_{\epsilon\to0}{1\over\epsilon^{d/2}}
\exp\left\{-{a\over2m\epsilon}({\boldsymbol q}_{i}^j-
{\boldsymbol q}_{i}^{j-1})^2\right\},
\label{deltarep2}
\end{eqnarray}
We keep $\epsilon$ finite until the end of the calculation.
Using this device, our formula for
the sum of planar diagrams is:
\begin{eqnarray}
T_{fi}&=&\lim_{\epsilon\to0}
\sum_{s_i^j=\pm1}\int DcDbD{\boldsymbol q}
 \exp\left\{\ln{\hat g}\sum_{ij}{1-s_i^js_i^{j-1}\over2}
-{d\over2}\ln\left(1+\rho\right)\sum_{i,j}P_i^j\right\}
\nonumber\\
&&\exp\left\{-{a\over2m}\sum_{i,j}
{({\boldsymbol q}_{i+1}^j-{\boldsymbol q}_{i}^{j})^2}
-{a\over2m\epsilon}\sum_{i,j}P_i^jP_i^{j-1}
{({\boldsymbol q}_{i}^j-{\boldsymbol q}_{i}^{j-1})^2}
\right\}\label{isingsumeps2}\\
&&\exp\left\{{a\over m}
\sum_{i,j}\left[A_{ij}{\boldsymbol b}^j_{i}{\boldsymbol c}^j_{i}
-B_{ij}{b}_{i}^{j}{c}_{i}^{j}
+C_{ij}({\boldsymbol b}_{i+1}^j-{\boldsymbol b}_i^j)
({\boldsymbol c}_{i+1}^j-{\boldsymbol c}_i^j)
-D_{ij}({b}_{i+1}^j-{b}_i^j)({c}_{i+1}^j-{c}_i^j)\right]\right\}\nonumber\\
A_{ij}&=&{1\over\epsilon}{P}_i^j{P}_i^{j-1}
+{P}_i^{j+1}{P}_i^j-{P}_i^{j-1}{P}_i^j{P}_i^{j+1}
+(1-P_i^j)(P_{i+1}^j+P_{i-1}^j)+\rho(1-P_i^j)P_{i-1}^{j-1}P_{i-1}^j\\
B_{ij}&=&(1-P_i^j)\left(P_{i+1}^jP_{i+1}^{j+1}
(1-P_{i+1}^{j-1})+
P_{i-1}^j{P_{i-1}^{j+1}(1-P_{i-1}^{j-1})}
+P_i^{j-1}P_i^{j-2}P_{i+1}^j\right)\\
C_{ij}&=&(1-P_i^j)(1-P_{i+1}^j)
\\
D_{ij}&=&(1-P_i^j)(1-P_{i+1}^j)P_i^{j-1}P_i^{j-2}
\end{eqnarray}
The first exponent in (\ref{isingsumeps2}) supplies a factor of
${\hat g}$ whenever a boundary is created or destroyed.
The second exponent includes the action $S_q$ for the free
propagator together with the exponent in the Gaussian representation
of the delta function that enforces Dirichlet boundary conditions
on the solid lines.
The first term of the third exponent incorporates the $\epsilon$ dependent
prefactor in the representation of the delta
function as a term in the ghost Lagrangian. 
The remaining terms contain $S_g$ together with
strategically placed spin projectors that arrange the
proper boundary conditions on the Grassmann variables
and supply appropriate $1/p^+$ factors needed at the
beginning or end of solid lines to ensure Lorentz invariance.
\section{Self Energy for $\Phi^3$}
\begin{figure}[htb]
\begin{center}
\psfrag{'k1'}{$k_1=k_0+k$}
\psfrag{'k0'}{$k_0$}
\psfrag{'l'}{$l$}
\psfrag{'M'}{$M$}
\includegraphics[width=6cm]{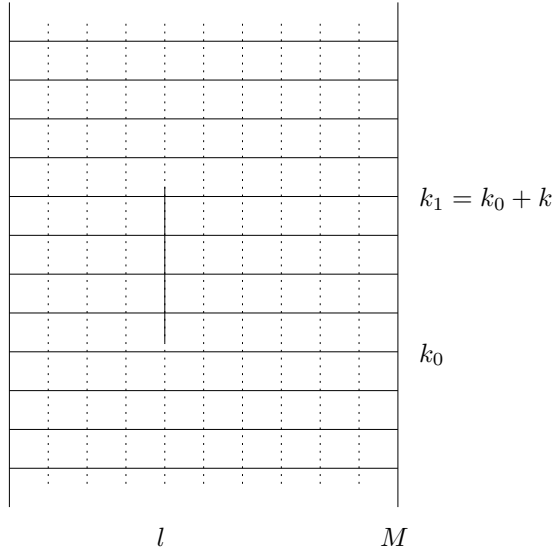}
\caption{Worldsheet lattice for self-energy}
\label{selfenergy}
\end{center}
\end{figure}

The worldsheet lattice for the one loop self energy is drawn in
Fig.~\ref{selfenergy}. 
The solid line segment in the
middle of the diagram is the internal boundary
that separates the two propagators of the
two gluon intermediate state. We take $\epsilon=0$ from the
beginning, so exact Dirichlet boundary conditions are imposed.
For this single diagram, the worldsheet path integral, 
(\ref{isingsumeps2}) with $\epsilon\to0$
and $\boldsymbol{q}_l^j=\boldsymbol{q}$ fixed on the
internal boundary, immediately reduces
to the usual light-cone Feynman rules. 
Then we evaluate the ${\boldsymbol q}$ integration
and take $N$ large:
\bea
T_{\rm SE} &=&\left({a\over2\pi m}\right)^{d/2}
\sum_{k_0,k,l}{{\hat g}^2\over Ml(M-l)}\int d\boldsymbol{q}
\exp\left\{{-Na({\boldsymbol p}^2+\mu^2)\over2mM}
-{kaM\over2ml(M-l)}\left(\boldsymbol{q}^2
+{\mu^2}{{M^2-l(M-l)\over M^2}}\right)\right\}
\nonumber\\
&\sim& N{{\hat g}^2\over M^2}\sum_{k,l}
\left[{l(M-l)\over M}\right]^{{d\over2}-1}{1\over k^{d/2}}
\exp\left\{-{ka\mu^2\over2m}\left({M\over l(M-l)}-{1\over M}\right)
\right\}
\eea
The factor of $N=T/a$ simply reflects time translational invariance,
and leads to the interpretation of its coefficient as $-a$
times a shift in energy, $-a\delta p^-=-a\delta\mu^2/2mM$.
Thus we have
\bea
\delta\mu^2=-{2m{\hat g}^2\over aM}\sum_{k,l}
\left[{l(M-l)\over M}\right]^{{d\over2}-1}{1\over k^{d/2}}
\exp\left\{-{ka\mu^2\over2m}\left({M\over l(M-l)}-{1\over M}\right)
\right\}.
\eea
We see that the worldsheet lattice has provided a cutoff
for the usual field theoretic ultraviolet divergences.
The removal of this cutoff in this quantity is simply taking
the limit $M\to\infty$, which is the limit of a continuous
worldsheet. More generally, for the
finite time transition amplitude,
the continuum limit is $M,N\to\infty$, with
$Na/Mm=T/p^+$ fixed.

We first observe that this limit is well defined as long as
$d<2$. The $k,l$ sums go to integrals over
continuous variables $t=(ka\mu^2/2m)\left({M/l(M-l)}-{1/M}\right)$
and $x=l/M$:
\bea
\sum_k{1\over k^{d/2}}&\to&
\int {dt\over t^{d/2}}
\left[{a\mu^2\over2m}\left({M\over l(M-l)}-{1\over M}\right)\right]^{(d-2)/2},
\qquad
{1\over M}\sum_l\to \int_0^1 dx\phantom{{1\over M}}\nonumber
\eea
Then we find
\bea
\delta\mu^2\to -\mu^2{\hat g}^2
\left[a\mu^2\over2m\right]^{(d-4)/2}
\Gamma\left(1-{d\over2}\right)\int_0^1dx(1-x(1-x))^{(d-2)/2}
+ O(M^{(d-2)/2}).\nonumber
\eea
\vskip12pt
\noindent Order by order in perturbation
theory, one could invoke
dimensional regularization, which defines divergent quantities
as the continuation in dimension from a region
where they are finite. Then we would
conclude from this formula
that the mass shift is Lorentz invariant.
Divergences in mass shift would appear as 
Lornetz invariant poles at integer $d\geq 2$,
and they could be covariantly absorbed in the input mass parameter
$\mu$.

However, dimensional regularization is pretty useless for
non-perturbative numerical work, because, in the presence of an
ultraviolet cutoff (necessary
for digitizing the problem), calculations at general $d$
contain non-covariant artifacts, which are
negligible only in sufficiently small dimensions. It is
essential in dimensional regularization to take
the continuum limit with $d$ in a range where the
quantity is finite {\it before} continuing to the
physical dimension, a procedure that is impossible on
a computer.
 
Indeed, the cutoff provided by the worldsheet lattice introduces
non-covariant artifacts simply because the cutoff $M=p^+/m$
is a component of momentum so all divergent terms
introduce frame dependence. We can see why this
happens by noting that the corrections to the continuum limit
for $d<2$ are of $O(M^{(d-2)/2})$ and obviously non-covariant.
As $d\to2$ these terms fall off more and more slowly
eventually becoming comparable to the Lorentz invariant term
and then for $d>2$ dominating it. Thus the worldsheet lattice by itself
will consistently produce Lorentz invariant results
only when divergences are completely absent, i.e. for $d<2$.

If we want to numerically analyze the system for $d\geq2$
without introducing an additional
{\it uv} cutoff, the worldsheet lattice must
be supplemented with explicit counter-terms that remove
the Lorentz violating artifacts introduced by divergences.
For example, at $d=2$ a logarithmic divergence in the self energy
appears as $\ln M =\ln(1/m) + \ln P^+$, and the non-covariance
is actually in the finite part.
There is the distinct possibility that the necessary
counter-terms are not local on the worldsheet, though it is relatively
easy to find local counter-terms that fix the problems in the
self energy at least for $d=2,4$.

To see this, take the interesting asymptotically free case of
$\Tr\Phi^3$ scalar field theory in $D=6$ space-time dimensions ($d=4$). 
Then in the self-energy one encounters the $\delta\to0$ limit of the quantity
\bea
&&\sum_{k=1}^\infty {e^{-k\delta}\over k^2}\to {\pi^2\over6}+\delta\ln\delta
+O(\delta^2),
\eea
and one finds for the mass shift
\bea
\delta\mu^2&\to& -{2m\over a}{\hat g}^2\sum_l x\left(1-x\right)
\left[{\pi^2\over6}+\delta_l\ln\delta_l\right]
\nonumber\\
 x&=&{l\over M},\qquad \delta_l={a\mu^2\over2Mm}{1-x(1-x)\over x(1-x)},
\nonumber
\eea
and then, as $M\to\infty$, the behavior
\bea
\delta\mu^2\to C_1 M +C_0^\prime\ln M +C_0.
\eea
Clearly the $C_1,C_0^\prime$ terms are non-covariant, and
they must be removed by counter-terms. This can be done
{\it locally} on the worldsheet. First note
that the $C_1$ term can be canceled by a constant 
energy shift, which can be interpreted as a worldsheet
boundary term (i.e. a perimeter cosmological constant).
Then one can devise an isolated 
up spin with ghost insertions to contribute 
a counter-term $\sum_l{1\over l}\sim\ln M$ to cancel $C_0^\prime$. 
The remaining $C_0$ term is just a covariant mass shift.
It is not at all clear, however, that
the counter-terms needed in vertex loop diagrams
can also be prescribed {\it locally} on the worldsheet.
One of the longstanding drawbacks of
standard lightcone gauge perturbation theory is 
the need for counter-terms that are not polynomials in the $p^+$'s, 
and there is no simple {\it a priori} principle for specifying them.
It is possible that worldsheet
locality provides such a principle.
If true, this would give an {\it a priori}
justification to apply the logic of string theory to help 
define and solve large $N_c$ QCD. Bardakci \cite{bardakci} has
stressed that $M^{+-}$ boost invariance
is precisely worldsheet scale invariance. Perhaps the
remaining  $M^{k-}$ Lorentz invariance
is the underlying physical reason for worldsheet conformal invariance.

In the following we take an alternative more systematic
approach, based on observations Glazek has made about controlling
Lorentz invariance in the light-cone formalism
\cite{glazek}. The troubles outlined above can be traced
to the way the worldsheet lattice cuts off the transverse
momentum integrals, $\Lambda_\perp \propto Mm/a$, so
the continuous worldsheet $M\to\infty$ has no {\it uv} cutoff.
This can be cured by introducing an $M$ independent
cutoff on transverse momentum which is held fixed as
$M\to\infty$. The theoretical drawback
is that it sacrifices Galilei invariance, a subgroup of
Lorentz group. However we shall find that the
problem of restoring this invariance is not severe.

The simplest way to implement an $M$ independent 
{\it uv} cutoff in transverse target space 
is to include a factor
$e^{-\delta\boldsymbol{q}_k^{j2}/2}$ 
in the world sheet path integrand whenever $(k,j)$ marks 
the beginning of a solid line.\footnote{This device
was first used in \cite{bardakcitimp} to facilitate
a mean field approximation \cite{bardakcitmean}
on the worldsheet.}
That is, we add terms $\delta(1-P_k^{j-1})P_k^j\boldsymbol{q}_k^{j2}/2$
to the worldsheet action. These terms
obviously violate Galilei invariance
(a part of Lorentz invariance), and
we must be careful that the invariance is restored after the
limit $\delta\to0$.

Let us now redo the self energy calculation with $\delta\neq0$. 
\bea
\delta\mu^2&=&-{2m{\hat g}^2\over a}\sum_{k,l}
{1\over l(M-l)}{1\over (kM/l(M-l)+m\delta/a)^{d/2}}
\nonumber\\
&&
\exp\left\{-{ka\mu^2\over2m}\left({M\over l(M-l)}-{1\over M}\right)
-{\delta\over2}k\left[{\boldsymbol{p}_0^2/l+\boldsymbol{p}_1^2/(M-l)
-(\boldsymbol{p}_1-\boldsymbol{p}_0)^2/M\over k/l+k/(M-l)+m\delta/a}
\right]\right\}.
\eea
Now we can safely take the continuum limit $M\to\infty$. Define
$T=kaM/l(M-l)m$ and $x= l/M$, we find
\bea
\delta\mu^2&=&-{2{\hat g}^2}\left({m\over a}\right)^{2-d/2}
\int_0^1 dx \int_0^\infty dT
{1\over (T+\delta)^{d/2}}
\nonumber\\
&&
\exp\left\{-{T\mu^2\over2}\left(1-x(1-x)\right)
-{\delta\over2}T\left[{(1-x)\boldsymbol{p}_0^2+x\boldsymbol{p}_1^2
-x(1-x)(\boldsymbol{p}_1-\boldsymbol{p}_0)^2\over T+\delta}
\right]\right\}.
\eea
The explicit dependence on the boundary values
of $\boldsymbol{q}$ reflects the violation of Galilei
boost invariance introduced by the cutoff $\delta$: this is the
price paid for regaining manifest longitudinal Lorentz boost 
invariance. Inspection of the formula shows that these
Galilei boost violations will disappear for $\delta\to0$
as long as $d<4$, i.e. in less that 6 space-time dimensions.
In this case, the divergences can be absorbed in a 
shift of $\mu^2$ consistently with Lorentz invariance
and with no counter-terms. 

We do want to study the 6 dimensional case, so we don't
quite escape the need for counter-terms. To study this issue for the
self energy, we set $d=4$ and analyze the $\delta\to0$
behavior of the mass shift.
\bea
\delta\mu^2&=&-{2{\hat g}^2}
\int_0^1 dx e^{\delta(\alpha(x)-\beta(x))}\int_\delta^\infty dT
{1\over T^2}\exp\left\{-T\alpha(x)+{\delta^2\over T}\beta(x)\right\}
\nonumber\\
&=&-2{\hat g}^2 \int_0^1 dx\left(
{1\over\delta}+\ln\delta -{\beta(x)\over2}
-\alpha^2(x)\int_0^\infty \ln T e^{-\alpha(x)T}\right) + O(\delta)
\nonumber\\
\alpha(x)&=&{\mu^2\over2}\left(1-x(1-x)\right) \\
\beta(x)&=&{1\over2}\left[{(1-x)\boldsymbol{p}_0^2+x\boldsymbol{p}_1^2
-x(1-x)(\boldsymbol{p}_1-\boldsymbol{p}_0)^2}
\right]={1\over2}\left[(1-x)\boldsymbol{p}_0+x\boldsymbol{p}_1\right]^2
\eea
In the $\delta\to0$ limit the non-covariant artifact resides
in the term ${\hat g}^2\int dx \beta(x)$, a
finite positive contribution to $\delta\mu^2$. The
corresponding contribution to the path integral is
of course 
\bea
-T\delta\mu^2/2p^+=-(aN/2mM){\hat g}^2\int dx \beta(x).
\label{noncovse}
\eea
All of the divergences can be covariantly
absorbed in a mass shift. But we still need to
design a worldsheet local
counter-term that removes this finite but non-covariant artifact.

To construct a suitable counter term we recall from \cite{thornsheet}
the generating formula for correlators of $\boldsymbol{q}_i^j$
on a fixed time slice $j$ of the worldsheet path integral
representation of the free propagator:
 \bea
\left\langle \exp\left\{\sum_{i=1}^{M-1} J_iq_i\right\}\right\rangle
&=&\exp\left\{{m\over 2a}\sum_i {i(M-i)\over M}J_i^2
+{m\over a}\sum_{i<j}{i(M-j)\over M}J_iJ_j\right.\nonumber\\
&&\hskip3cm\left.+{q_M\over M}\sum_i iJ_i
+{q_0\over M}\sum_i(M-i)J_i\right\}\label{qgen}
\eea
Differentiating (\ref{qgen}) twice with respect to
$J_i$ and setting all $J=0$, we find
\bea
\langle {\boldsymbol q}_l^{j2}\rangle
&=&\left[{\boldsymbol{q}_M\over M}l
+{\boldsymbol{q}_0\over M}(M-l)\right]^2+{m\over a}{l(M-l)\over M}\\
{1\over M}\sum_{l=1}^{M-1}\langle {\boldsymbol q}_l^{j2}\rangle&=&
\int_0^1 dx\left[{\boldsymbol{p}_1}x
+{\boldsymbol{p}_0}(1-x)\right]^2+{mM\over6a} +O\left({1\over M}\right)\\
&=&2\int_0^1 dx \beta(x)+{mM\over6a} +O\left({1\over M}\right)
\eea
where in the last line we have used the boundary conditions 
$\bfs{q}_0=\boldsymbol{p}_0$, $\boldsymbol{q}_M=\boldsymbol{p}_1$.

Referring to (\ref{noncovse}), we see that we can represent the
necessary counter-term as
\bea
(aN/2mM){\hat g}^2\int dx \beta(x)&=&{a\over2m}\left[
{N\over2M^2}\sum_{l=1}^{M-1}\langle {\boldsymbol q}_l^{j2}\rangle
-{mN\over12a}\right]
=\sum_{j=1}^N\sum_{l=1}^{M-1}
\left\langle {\hat g}^2{a\over4mM^2}{\boldsymbol q}_l^{j2}\right\rangle
-{\hat g}^2{N\over24}.
\label{noncovct}
\eea
After representing the $1/M^2$ in the summand by local
modifications of the ghost action near the point $(l,j)$, the
first term can be seen as a sum over all locations of a
{\it local} world sheet insertion. The last term is
precisely of the right structure to be absorbed in a
boundary perimeter term (boundary ``cosmological constant'').
We have already seen in \cite{thorntfisheet} that such
a perimeter term is needed to properly include a mass
for the scalar field in the worldsheet description, so
it is not surprising that in the process of mass renormalization
we should be required to adjust its value to make the final
answer covariant.

Summarizing, we have found that if we use our new {\it uv} cutoff
$\delta$, then for $D<6$ the mass shift shows no 
non-covariant artifacts, and the divergence 
(for $4\leq D<6$) can be covariantly
absorbed in $\mu^2$. For $D=6$ there is a finite non-covariant
artifact in the mass shift which can be canceled by
a worldsheet local counter-term together with an
adjustment of the value of the boundary cosmological constant.
The remaining ultraviolet divergences are covariant and
can be absorbed in $\mu^2$.

\section{Wave function renormalization}
Before moving on to the three point vertex we need to
analyze wave function renormalization, which though
finite for $D<6$ will show log divergences at $D=6$,
which will contribute to the renormalization
of the coupling ${\hat g}$. For this it is convenient to
work in energy space by defining
\bea
T(E)=\sum_{N=1}^\infty e^{aEN}T_N
\eea
where $T_N$ is the amplitude for evolution through
$N$ time steps. Then the free gluon propagator
is simply
\bea
\Delta_0(p^2)=\sum_{N=1}^\infty \exp\left\{
(aE-\lambda)N-Na{({\bfs p}_1-{\bfs p}_0)^2+\mu^2\over2mM}\right\}
={1\over e^{a(p^2+\mu^2)/2mM+\lambda}-1}
\eea
where we have defined the off-shell four momentum 
$p=({\bfs p},p^+,p^-)=({\bfs p}_1-{\bfs p}_0, mM,E)$. We have
also included a boundary cosmological constant $\lambda=O({\hat g}^2)$
which, as we have seen, will be necessary to cancel
non-covariant artifacts in loop diagrams.

Now we include up to one loop corrections to the full propagator
\bea
\Delta(p^2)&=&\Delta_0(p^2)\left\{1+\Delta_0(p^2)(\Pi(p^2)+\Pi_{\rm C.T.})
+O({\hat g}^4)\right\}\\
\Pi(p^2)&=&{{\hat g}^2\over M}\sum_{k,l}
{1\over l(M-l)}{1\over (kM/l(M-l)+m\delta/a)^2}
\exp\left\{-{ka\over2m}{p^2+\mu^2\over M}-2k\lambda\right\}
\nonumber\\
&&
\exp\left\{
-{ka\mu^2\over2m}\left({M\over l(M-l)}-{1\over M}\right)
-{\delta\over2}k\left[{\boldsymbol{p}_0^2/l+\boldsymbol{p}_1^2/(M-l)
-(\boldsymbol{p}_1-\boldsymbol{p}_0)^2/M\over k/l+k/(M-l)+m\delta/a}
\right]\right\}.
\label{Delta}
\eea
Since both the mass shift $\delta\mu^2$ and $\lambda$ are
of order $O({\hat g}^2)$, we may, to this order, 
replace $\mu$ by its physical
value and drop $\lambda$ in the expression for $\Pi$.
We express $\Delta_0$ in terms of the physical mass
$\mu^2_{ph}=\mu^2+\delta\mu^2$, expand it to first order
in ${\hat g}^2$, and plug into (\ref{Delta})
\bea
\Delta_0(p^2)&=&{\hat\Delta}_0(p^2)
\left[1+{\hat\Delta}_0(p^2)\left({e^{a(p^2+\mu_{ph}^2)/2mM}}
\left({a\delta\mu^2\over2mM}-\lambda\right)+\Pi(p^2)+\Pi_{C.T.}\right)
+O({\hat g}^4)\right]\\
{\hat\Delta}_0(p^2)&=&{1\over e^{a(p^2+\mu_{ph}^2)/2mM}-1}
\eea
The physical mass is of course defined by the requirement
\bea
\Pi(-\mu_{ph}^2)+\Pi_{C.T.}
+\left({a\delta\mu^2\over2mM}-\lambda\right)+O({\hat g}^4)
=0,
\eea
which reproduces the
covariant mass shift already discussed.
When this condition is met the $p^2\to-m_{ph}^2$ limit
of the quantity in square brackets exists and is equal
to the wave function renormalization constant $Z$:
\bea
Z&=&\lim_{ p^2\to-m_{ph}^2}\left[1+{\hat\Delta_0}(p^2)
(p^2+\mu_{ph}^2)\left\{{a\over2mM}\left({a\delta\mu^2\over2mM}-\lambda\right)
+\Pi^\prime(-\mu_{ph}^2)\right\}\right]\\
&=& 1+{a\delta\mu^2\over2mM}-\lambda
+{2mM\over a}\Pi^\prime(-\mu_{ph}^2)
\eea 
The discussion so far has retained both the worldsheet lattice cutoffs and
the {\it uv} cutoff $\delta$. Now we simplify the expression for $Z$
by taking the worldsheet continuum limit $M\to\infty$
holding $\delta$ fixed:
\bea
Z&{}_{{\displaystyle\to}\atop{\scriptstyle M\to\infty}}&
1-\lambda-{\hat g}^2\int_0^1 dx x(1-x)\int_0^\infty
{TdT\over(T+\delta)^2}\exp\left\{-\alpha(x)T-{\delta T\over T+\delta}\beta(x)
\right\}\\
&{}_{{\displaystyle\sim}\atop{\scriptstyle \delta\to0}}&
1+{{\hat g}^2\over6}\ln(\delta\mu^2)-\lambda
+{\hat g}^2\int_0^1 dx x(1-x)\left(1-\Gamma^\prime(1)+\ln(\alpha(x)/\mu^2)
\right)
\eea
where in the last line we see the log divergence as $\delta\to0$.
From our earlier considerations, we know that the value of
$\lambda$ to this order should be $\lambda={\hat g}^2/24$.

\section{The triangle graph and coupling renormalization}
\begin{figure}[ht]
\begin{center}
\psfrag{'k2'}{$k_2$}
\psfrag{'k1'}{$k_1$}
\psfrag{'k0'}{$k_0$}
\psfrag{'l'}{$l$}
\psfrag{'M1'}{$M_1$}
\psfrag{'M3'}{$M$}
\psfrag{'p'}{$\hskip-.5cm\bfs{q}-\bfs{p}_0$}
\psfrag{'p1'}{$\bfs{p}_1-\bfs{p}_0$}
\psfrag{'p2'}{$\bfs{p}_2-\bfs{p}_1$}
\psfrag{'p3'}{$\bfs{p}_2-\bfs{p}_0$}
\includegraphics[width=12cm]{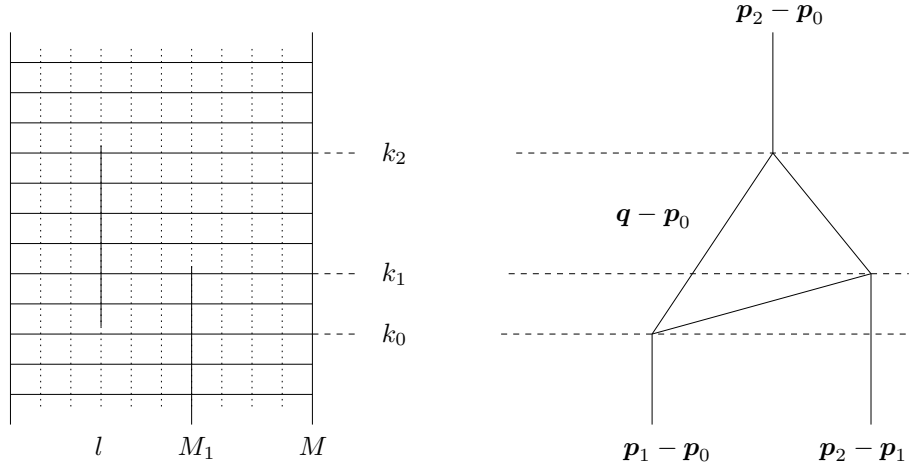}
\end{center}
\caption{Worldsheet for the triangle diagram shown on the right.}
\label{triangle}
\end{figure}

We first evaluate the 1PIR one loop correction
to the cubic vertex shown in Fig.~\ref{triangle}, 
which is finite for $d<4$. We 
do the calculation in the presence of the {\it uv} cutoff
$\delta$ introduced in the last section. There are two
relevant kinematic configurations in which the 
spatial location of the loop $l$ is in the range $0<l<M_1$
and $M_1<l<M$ respectively. We work out the first case,
depicted in the figure,
in great detail and then briefly discuss the second case.

We find for $0<l<M_1$, introducing an off shell energy $E=E_1+E_2$
\bea
G^1_3&=&\left({a\over2\pi m}\right)^{d/2}
{{\hat g}^3\over M}\sum\int d{\boldsymbol q}{1\over l(M_1-l)(M-l)}
\exp\left\{-{(N-k_2)a\over2m}
\left[{({\boldsymbol p}_2-{\boldsymbol p}_0)^2+\mu^2\over
M}\right]\right\}\nonumber\\
&&\exp\left\{-{(k_2-k_1)a\over2m}
\left[{({\boldsymbol q}-{\boldsymbol p}_0)^2+\mu^2\over
l}+{({\boldsymbol p}_2-{\boldsymbol q})^2+\mu^2\over
M-l}\right]-{\delta\over2}{\boldsymbol q}^2\right\}\nonumber\\
&&\exp\left\{-{(k_1-k_0)a\over2m}
\left[{({\boldsymbol q}-{\boldsymbol p}_0)^2+\mu^2\over
l}+{({\boldsymbol p}_1-{\boldsymbol q})^2+\mu^2\over
M_1-l}+{({\boldsymbol p}_2-{\boldsymbol p}_1)^2+\mu^2\over
M-M_1}\right]\right\}\nonumber\\
&&\exp\left\{-{k_0a\over2m}
\left[{({\boldsymbol p}_1-{\boldsymbol p}_0)^2+\mu^2\over
M_1}+{({\boldsymbol p}_2-{\boldsymbol p}_1)^2+\mu^2\over
M-M_1}\right]+aNE-ak^\prime_0E_2\right\}\\
&=&\left({a\over m}\right)^{d/2}
{{\hat g}^3\over M}\sum{1\over l(M_1-l)(M-l)}
{\Delta_0(Q_1^2)\Delta_0(Q_2^2)\Delta_0(Q_3^2)
\over(T_1+T_2+T_3+\delta)^{d/2}}\exp\left\{-{\mu^2\over2}(T_1+T_2+T_3)
\right\}\nonumber\\
&&\exp\left\{
-{T_1T_3Q^2_1+T_1T_2Q^2_2+T_2T_3Q_3^2\over 2(T_1+T_2+T_3)}-{\delta\over2}
\left[{({\boldsymbol p}_0T_3+{\boldsymbol p}_1T_1
+{\boldsymbol p}_2T_2)^2\over
(T_1+T_2+T_3+\delta)(T_1+T_2+T_3)}\right]\right\}
\eea
In this formula we have introduced the $T_i$ defined by
\bea
T_1={a\over m}{k_1-k_0\over M_1-l},\qquad T_2={a\over m}{k_2-k_1\over M-l},
\qquad T_3={a\over m}{k_2-k_0\over l}.
\eea
the integers $N_1=k_1-k_0$, $N_2=k_2-k_1$  range independently
over the positive integers. We have also introduced the
off shell $d+2$ momenta
\bea
Q_1=({\boldsymbol p}_1-{\boldsymbol p}_0,mM_1,E-E_2),\qquad 
Q_2=({\boldsymbol p}_2-{\boldsymbol p}_1,m(M-M_1),E_2),\qquad 
Q_3=({\boldsymbol p}_2-{\boldsymbol p}_0, mM, E). 
\eea
The sums over $N-k_2$,
$k_0$, and $k_1-k_0^\prime$ just produce the
external leg propagators $\Delta_0(Q_i^2)$. The integer $l$ takes all values
$0<l<M_1$. 

In the worldsheet continuum limit the sums over
$N_1,N_2,l$ will be replaced by integrals over $T_1,T_2,T_3$,
so we shall need the Jacobian
\bea
{\partial(T_1,T_2,T_3)\over\partial(N_1,N_2,l)}=\left({a\over m}\right)^2
{T_1+T_2+T_3\over l(M-l)(M_1-l)}.
\eea 
The full range $0<T_i<\infty$ is covered in the
continuum limit when the result $G^2_3$ of the calculation with
$M_1<l<M$ is combined with the one discussed above.
We then obtain for the worldsheet continuum limit with
fixed $\delta$ of the sum of both diagrams:
\bea
G^1_3+G^2_3&=&\Delta_0(Q_1^2)\Delta_0(Q_2^2)\Delta_0(Q_3^2)
{{\hat g}^3\over M}\left({a\over m}\right)^{(d-4)/2}
\int_0^\infty {dT_1dT_2dT_3\over T_1+T_2+T_3}
{(T_1+T_2+T_3+\delta)^{-d/2}}
\nonumber\\
&&\hskip-1in\exp\left\{-{\mu^2\over2}(T_1+T_2+T_3)
-{T_1T_3Q^2_1+T_1T_2Q^2_2+T_2T_3Q_3^2\over 2(T_1+T_2+T_3)}-{\delta\over2}
\left[{({\boldsymbol p}_0T_3+{\boldsymbol p}_1T_1
+{\boldsymbol p}_2T_2)^2\over
(T_1+T_2+T_3+\delta)(T_1+T_2+T_3)}\right]\right\}
\eea
We see explicitly that the $\delta\to0$ limit is finite
for $d<4$. For $d=4$ (6 space-time dimensions), the integral
is only logarithmically divergent in this limit so it is safe to set 
$\delta =0$ in the exponent for all $d\leq4$. Adding the
tree contribution, we see that up to one loop the 1PIR
three vertex is as $\delta\to0$ just the tree value times the factor
\bea
1+{{\hat g}^2}\left({a\over m}\right)^{(d-4)/2}
\int_0^\infty {dT_1dT_2dT_3\over (T_1+T_2+T_3)^{1+d/2}}
\exp\left\{-{\mu^2\over2}(T_1+T_2+T_3)
-{T_1T_3Q^2_1+T_1T_2Q^2_2+T_2T_3Q_3^2\over 2(T_1+T_2+T_3)}\right\}
\eea
for $d<4$. For $d=4$, we extract the log divergence by breaking the
integration domain into a region with $T_1+T_2+T_3>\epsilon$ 
for which we may set $\delta=0$ and a region $T_1+T_2+T_3<\epsilon$
for which we may drop the exponent and then evaluate it explicitly:
\bea
\int_{T_1+T_2+T_3<\epsilon}{dT_1dT_2dT_3\over (T_1+T_2+T_3)
(T_1+T_2+T_3+\delta)^2}&=&{1\over2} \int_0^\epsilon {TdT\over(T+\delta)^2}
\sim{1\over2}\left(\ln{\epsilon\over\delta}-1\right)\nonumber
\eea
Then the modification factor for $d=4$ can be written
\bea
&\sim&1+{{\hat g}^2\over2}\left(\ln{\epsilon\over\delta}-1\right)+
{{\hat g}^2}\int_{T_1+T_2+T_3>\epsilon} {dT_1dT_2dT_3\over (T_1+T_2+T_3)^3}
\nonumber\\
&&\exp\left\{-{\mu^2\over2}(T_1+T_2+T_3)
-{T_1T_3Q^2_1+T_1T_2Q^2_2+T_2T_3Q_3^2\over 2(T_1+T_2+T_3)}\right\}
\eea
Incorporating the wave function renormalization factors $Z^{3/2}$,
we see that the divergence can be absorbed in a renormalized
coupling
\bea
{\hat g}_{r}&=&{\hat g}\left(1+{{\hat g}^2\over2}\ln{1\over\mu^2\delta}
+{3\over2}{{\hat g}^2\over6}\ln{\mu^2\delta}\right)
={\hat g}\left(1+{{\hat g}^2\over4}\ln{1\over\mu^2\delta}\right)
\eea
Recall that ${\hat g}^2=g^2/64\pi^3$, where $g$ is the conventionally
defined coupling. In terms of it, the renormalization reads
\bea
{g}_{r}&=&{g}\left(1+{{g}^2\over256\pi^3}\ln{1\over\mu^2\delta}\right)
\eea
and the Callan-Symanzik beta function is
\bea
\beta(g_r)\equiv\mu{dg_{r}\over d\mu}=-{g_r^3\over128\pi^3}+O(g_r^5) 
\label{largenbeta}
\eea
To compare to the standard result, remember that this result
is strictly the $N_c\to\infty$ limit, and $g$ is $\sqrt{N_c}$
times the conventional coupling. 
At finite $N_c$ one can decompose
$\Phi$ into adjoint and singlet components. 
Then one finds different renormalizations
for the $Adj^3$, $Adj^2 S$ and $S^3$ vertices. The corresponding
beta functions are (\ref{largenbeta}), (\ref{largenbeta}) times 8/3,
and (\ref{largenbeta}) times 6, respectively. Then the $N_c=1$
beta function is to be compared to the one for the $S^3$ vertex.
The field $\Phi$ for the case $N_c=1$ is just a single hermitian scalar field.
But with our definition the, cubic term goes to $g\Phi^3/3$
instead of the customary  $g\Phi^3/3!$; after taking this
difference into account, which means multiplying our beta
function by $1/4$, we get the known result for $N_c=1$,
with $-3/256\pi^3$ multiplying the customary coupling cubed.

\section{Conclusion}
The worldsheet ``template'' for summing planar diagrams has
been set up for a whole range of interesting theories,
including QCD and supersymmetric gauge theories. In this talk
I have shown in detail how the field theoretic
renormalization program plays out on the worldsheet
for theories with scalar cubic couplings.
We have given in this case
a local worldsheet description of the
counter-terms necessary for Lorentz invariance
for space-time dimensions $D\leq 6$. More generally it is
hoped that the principle of worldsheet locality will assist 
the renormalization program for gauge theories in light-cone
gauge. Because field theoretic locality
is not manifest in this gauge it is of no direct use in
restricting counter-terms. The new worldsheet locality,
if it survives the regulation of {\it uv} divergences,
will be manifest and will therefore provide a new
principle for classifying counter-terms. The main
obstacle still to be overcome is the worldsheet
regulation of the $p^+=0$ singularities that occur
in gauge theories in light-cone gauge.

The eventual goal of the worldsheet description of field theory
is to bring the powerful techniques of string theory to bear
on the problem of quark confinement in QCD.\footnote{This goal is of
course shared by practitioners of the AdS/CFT
correspondence \cite{maldacena,klebanovs,polchinskis,berensteinmn}.} 
There is a remote chance that
it will enable a completely analytic understanding of this important problem.
But even if this is not possible, a numerical
attack on the worldsheet formulation of
the problem may offer insights complementary to those
provided by lattice gauge theory. In particular, the
two-dimensionality of the worldsheet lattice promises to
bring new efficiencies to numerical spectrum calculations,
perhaps allowing a closer approach to the continuum answers.

\noindent\underline{ Acknowledgments}: 
I am grateful to K. Bardakci and S. Glazek 
for valuable discussions. This research
was supported in part by the Department
of Energy under Grant No. DE-FG02-97ER-41029.


\begin{thebibliography}{1}
\bibitem{bardakcit}
K.~Bardakci and C.~B.~Thorn, Nucl. Phys. {\bf B626} (2002) 287, hep-th/0110301.
\bibitem{thooftlargen}
G. 't Hooft, {\sl Nucl. Phys.} {\bf B72} (1974) 461.
\bibitem{thornsheet}
C.~B.~Thorn,
Nucl.\ Phys.\ B {\bf 637} (2002) 272
[arXiv:hep-th/0203167].
\bibitem{gudmundssontt}
S.~Gudmundsson, C.~B.~Thorn, and T.~A.~Tran
Nucl.\ Phys.\ B {\bf 649} (2002) 3, 
[arXiv:hep-th/0209102].
\bibitem{thorndurham} C.~B.~Thorn, [arXiv:hep-th/0310121].
\bibitem{goddardgrt}
P. Goddard, J. Goldstone, C. Rebbi, and C. B. Thorn, { Nucl. Phys.} {\bf
  B56} (1973) 109.
\bibitem{gilest}
R. Giles and C. B. Thorn, {Phys. Rev.} {\bf D16} (1977) 366;
C. B. Thorn, {Phys. Lett.} {\bf 70B} (1977) 85; 
{Phys. Rev.} {\bf D17} (1978) 1073;
K. Bering, J. S. Rozowsky and C. B. Thorn, 
{Phys. Rev.} {\bf D61} (2000) 045007, hep-th/9909141.
\bibitem{bardakcitimp}
K.~Bardakci and C.~B.~Thorn,
Nucl. Phys. B {\bf 661} (2003) 235, [arXiv:hep-th/0212254].
\bibitem{bardakcitmean}
K.~Bardakci and C.~B.~Thorn,
Nucl. Phys. B {\bf 652} (2003) 196,[arXiv:hep-th/0206205].
\bibitem{bardakci}
K.~Bardakci,
``Self consistent field method for planar phi**3 theory,''
arXiv:hep-th/0308197.
\bibitem{glazek}
S.~D.~Glazek,
Phys.\ Rev.\ D {\bf 60} (1999) 105030 [arXiv:hep-th/9904029]; 
Phys.\ Rev.\ D {\bf 63} (2001) 116006 [arXiv:hep-th/0012012]; 
Phys.\ Rev.\ D {\bf 66} (2002) 016001 [arXiv:hep-th/0204171]; 
arXiv:hep-th/0307064.
\bibitem{thorntfisheet} 
C.~B.~Thorn and T.~A.~Tran,
arXiv:hep-th/0307203.
\bibitem{maldacena}
J. M. Maldacena, {\sl Adv. Theor. Math. Phys.} {\bf 2} (1998) 231-252,
  hep-th/9711200.
\bibitem{klebanovs}
I.~R.~Klebanov and M.~J.~Strassler,
JHEP {\bf 0008} (2000) 052
[arXiv:hep-th/0007191].
\bibitem{polchinskis}
J.~Polchinski and M.~J.~Strassler,
arXiv:hep-th/0003136;
Phys.\ Rev.\ Lett.\  {\bf 88} (2002) 031601
[arXiv:hep-th/0109174];
JHEP {\bf 0305} (2003) 012
[arXiv:hep-th/0209211].
\bibitem{berensteinmn}
D.~Berenstein, J.~M.~Maldacena and H.~Nastase,
JHEP {\bf 0204} (2002) 013
[arXiv:hep-th/0202021].
\end{thebibliography}
\end{document}